\documentclass[sigconf]{acmart}
\setcopyright{none}
\renewcommand\footnotetextcopyrightpermission[1]{}
\makeatletter
\let\@authorsaddresses\@empty
\makeatother
\AtBeginDocument{%
  }

\usepackage{multirow}
\usepackage{booktabs}
\usepackage{siunitx}
\usepackage{dcolumn}
\usepackage{amsmath}

\begin{document}

\title{Can LLMs Address Mental Health Questions? \\ A Comparison with Human Therapists}

\author{Synthia Wang}
\affiliation{
  \institution{University of Chicago}
  \city{Chicago}
  \state{Illinois}
  \country{USA}
}
\author{Yuwei Cheng}
\affiliation{
  \institution{University of Chicago}
  \city{Chicago}
  \state{Illinois}
  \country{USA}
}
\author{Austin Song}
\affiliation{
  \institution{University of Virginia}
  \city{Charlottesville}
  \state{Virginina}
  \country{USA}
}
\author{Sarah Keedy}
\affiliation{
  \institution{University of Chicago}
  \city{Chicago}
  \state{Illinois}
  \country{USA}
}
\author{Marc Berman}
\affiliation{
  \institution{University of Chicago}
  \city{Chicago}
  \state{Illinois}
  \country{USA}
}
\author{Nick Feamster}
\affiliation{
  \institution{University of Chicago}
  \city{Chicago}
  \state{Illinois}
  \country{USA}
}

\begin{abstract}
Limited access to mental health care has motivated the use of digital tools and conversational agents powered by large language models (LLMs), yet their quality and reception remain unclear. We present a study comparing therapist-written responses to those generated by ChatGPT, Gemini, and Llama for real patient questions. Text analysis showed that LLMs produced longer, more readable, and lexically richer responses with a more positive tone, while therapist responses were more often written in the first person. In a survey with 150 users and 23 licensed therapists, participants rated LLM responses as clearer, more respectful, and more supportive than therapist-written answers. Yet, both groups of participants expressed a stronger preference for human therapist support. These findings highlight the promise and limitations of LLMs in mental health, underscoring the need for designs that balance their communicative strengths with concerns of trust, privacy, and accountability.
\end{abstract}

\maketitle
\pagestyle{plain}
\section{Introduction}
The global demand for mental health support continues to increase at an unprecedented pace. In the United States, over 20\% of adults live with a diagnosable mental illness each year~\cite{nimh2023mental}, and major depression is projected to become the leading cause of disease worldwide by 2030~\cite{patel2018depression}. At the same time, many individuals lack timely or affordable access to care due to shortages of licensed professionals, long waitlists, and other structural barriers. Access is further limited by social factors; stigma surrounding mental health varies across cultures, countries, and even local contexts (e.g., urban versus rural settings), preventing many from seeking professional help.

On the other hand, large language models (LLMs) have rapidly entered everyday life in recent years. Recent surveys suggest that nearly half of U.S. adults with ongoing mental health conditions have experimented with LLMs for therapeutic support \cite{forerunnerventures2025}, while roughly a third of people in the U.S. and U.K. now use generative AI daily for purposes ranging from productivity to health advice \cite{techradar2025, nypost2025}. These trends highlight both the increasing availability of AI tools and the growing reliance on alternative sources of support when traditional mental health resources are limited. Together, these developments highlight the urgent need to investigate not only how LLMs can complement existing mental health resources but also the quality, effectiveness, and limitations of the support they can provide in real-world contexts.

While conversational agents are not new in this space, with structured tools such as Woebot~\cite{fitzpatrick2017delivering} and Wysa~\cite{inkster2018empathy} having long demonstrated feasibility and engagement in delivering cognitive behavioral therapy, the advent of LLMs has transformed expectations around accessibility, personalization, and scale. Recent work has explored their ability to generate treatment plans~\cite{james2023towards}, support journaling~\cite{kim2024mindfuldiary}, assist with cognitive restructuring~\cite{wang2025evaluating}, and provide clinically relevant explanations~\cite{krishnamoorthy2025demystifying, yang2023towards}. Yet LLMs also introduce new risks around hallucinations (i.e., providing inaccurate information), lack of cultural sensitivity, and ethical concerns around safety, privacy, accountability, and emotional dependency~\cite{cabrera2023ethical, guo2024large, lawrence2024opportunities, ma2024understanding}. Although prior studies indicate that LLMs are capable of classifying mental health issues and providing psychiatric knowledge~\cite{hanss2025assessing, lamichhane2023evaluation}, there remains little systematic evidence comparing LLM responses with those written by licensed professionals. Even fewer are evaluations that consider the perspectives of both end users and therapists, which allows for a more complete understanding of trust, acceptability, and practical utility, all of which are essential for responsibly integrating LLMs into mental health care. Such studies can guide the design and deployment of LLM-driven interventions that are both effective and ethically sound.

To address these gaps, our study explores three research questions:
\begin{itemize}
    \item RQ1: What are the differences between LLM-generated responses to mental health questions compared to licensed therapists’ responses?
    \item RQ2: How do users perceive LLM-generated responses to mental health questions compared to licensed therapists’ responses?
    \item RQ3: How do licensed therapists perceive LLM-generated responses to mental health questions compared to licensed therapists’ responses?
\end{itemize}

To answer our research questions, we examined how users and licensed therapists perceived LLM-generated versus therapist-written answers to mental health-related questions available in the Counsel Chat dataset~\cite{counselchat}. Using it, we paired its therapist-written answers with responses to the same questions generated by three widely used LLMs, namely ChatGPT, Gemini, and Llama. In our survey study, 150 U.S.-based users and 23 licensed therapists rated these answers on clarity, empathy, respect, and overall quality, while therapists also assessed professional acceptability. Complementing these perceptual evaluations, we conducted a text analysis to examine linguistic and stylistic differences, including readability, vocabulary diversity, sentiment, and hedging, between the LLM responses and the therapists' responses. 

Our findings reveal that general participants and therapists consistently rated LLM responses higher than therapist responses on dimensions such as clarity, encouragement, and respectfulness. Linguistic analyses revealed that LLM responses were generally longer, lexically richer, and exhibited a more neutral or positive tone, while therapist-written responses were more readable and frequently employed personal framing. We also found neither group was particularly good at distinguishing LLM-generated answers from human-written answers, which fits well with our observation that, despite the observed preference for the LLM responses, this preference did not translate into greater trust in the models. 74\% of our participants reported that, if they needed support, they would prefer to seek help from a human therapist than from an LLM. In addition, therapists expressed reluctance to recommend LLMs beyond providing general information, frequently citing concerns about incorrect or potentially harmful content.

Taken together, these results highlight both the promise, pitfalls, and the tension of deploying LLMs in sensitive domains such as mental health. On the one hand, LLMs demonstrate communicative competence that rivals professional responses in perceived quality. On the other hand, their lack of accountability, contextual judgment, and professional oversight continues to limit their acceptability as a trusted source of care. It also remains unknown how our participant ratings would translate to real-world efficacy in a clinical setting. However, by providing one of the first systematic comparisons of LLM-generated and therapist-written responses to real mental health questions, our work contributes empirical evidence and design insights to ongoing debates about the role of generative AI in augmenting or supplementing existing therapies. Importantly, our findings do not imply that LLMs can substitute for professional mental health care. Instead, this work demonstrates a proof-of-concept framework for comparing LLM and therapist responses, providing insights that could help guide the development of scalable, AI-assisted mental health support.
\section{Background}
Our work builds on prior research in conversational agents for mental health as well as the emerging use of large language models in therapeutic contexts. In this section, we situate our comparative evaluation within the broader landscape of digital mental health interventions, highlighting both their opportunities and their risks.

\subsection{Conversational Agents in Mental Health}
Digital conversational agents (CAs) have been under development for many years as a means of providing accessible mental health support. Well-developed chatbots such as Woebot have been studied extensively, demonstrating that CAs can provide structured cognitive behavioral therapy (CBT) in ways that are feasible, engaging, and effective~\cite{fitzpatrick2017delivering}. Similarly, the Wysa app has shown promise in supporting individuals with self-reported depressive symptoms, highlighting the potential for digital tools to reach populations that may not have access to traditional therapy~\cite{inkster2018empathy}. While therapeutic bonds have traditionally been viewed as uniquely human, evidence indicates that digital interventions can foster meaningful therapeutic relationships with users, supporting emotional engagement and self-disclosure~\cite{darcy2021evidence,lee2020hear}. Features such as chatbot self-disclosure appear particularly effective in promoting reciprocal disclosure, enhancing perceived intimacy, and improving enjoyment~\cite{lee2020hear}. Beyond individual engagement, chatbot-based social contact has been identified as a promising approach for reducing mental illness stigma, and potentially complementing broader public health efforts~\cite{lee2023exploring,reategui2025llm}.

The effectiveness of CAs depends, in part, on their design characteristics and integration context. Multimodal CAs, generative AI-based systems, and those embedded in mobile or instant messaging platforms tend to achieve higher engagement and adherence, particularly among clinical, subclinical, and elderly populations~\cite{li2023systematic}. However, observed improvements in overall psychological well-being are modest, highlighting the need to explore underlying mechanisms of engagement and clinical impact. Trust, usability, and accessibility are critical components of user experience, and partly intersect with components of effective psychotherapy, which itself remains to be fully understood~\cite{carey2020identifying}. Studies reveal that participants often trust chatbots for tasks such as logging mood or gratitude, providing advice, and offering guidance, while concerns remain regarding data privacy and the storage of personal information~\cite{boyd2022usability,sweeney2021can,bird2023generative,haque2023overview}. Other assessments also indicate that chatbots can produce measurements consistent with traditional methods, although they may require greater effort and time from users~\cite{schick2022validity}. At the same time, CAs face challenges in delivering nuanced emotional support, maintaining consistent interactions, and avoiding overreliance~\cite{denecke2021artificial,pham2022artificial}. Despite their potential for cost-effective and stigma-free interventions, ethical and legal considerations, including responsibility for incorrect advice and privacy of protected health information, remain crucial for their responsible deployment~\cite{pham2022artificial}.

\subsection{LLMs in Mental Health}
The emergence of LLMs has introduced new possibilities for personalized, scalable, and explainable mental health support. LLM-based applications have been explored experimentally across multiple domains, including journaling~\cite{kim2024mindfuldiary}, cognitive restructuring~\cite{wang2025evaluating}, mental health classification~\cite{lamichhane2023evaluation,xu2024mental}, and treatment plan generation~\cite{james2023towards}. LLMs demonstrate strong in-context learning capabilities and, when combined with few-shot prompting and emotional cue integration, can approach human-level performance in generating clinically relevant explanations~\cite{yang2023towards,krishnamoorthy2025demystifying}. Frameworks such as MentalBlend illustrate how cognitive-behavioral, dialectical behavior, person-centered, and reality therapy principles can be embedded within LLM responses to produce outputs aligned with professional standards~\cite{gu2024mentalblend}. Edge-deployed systems like MindGuard integrate subjective ecological assessments with sensor data to deliver stigma-free, personalized screening and interventions, achieving performance comparable to or exceeding large general-purpose models~\cite{ji2024mindguard}. LLMs have also been used to simplify complex psychiatric information, supporting both patient understanding and clinician decision-making~\cite{krishnamoorthy2025demystifying}.

Evaluations of LLM-based interventions suggest that these models can offer high-quality support in areas such as diagnostics, treatment planning, and psychoeducation~\cite{levkovich2025evaluating,li2025counselbench,hanss2025assessing,kuhlmeier2025combining,reategui2025llm,hua2024large}. Studies indicate that LLMs have promising capability for mental health classification tasks and can provide reliable general psychiatric knowledge, with repeated prompting and careful design improving response confidence and explainability~\cite{hanss2025assessing,lamichhane2023evaluation}. However, user experience and clinical effectiveness are shaped not only by the LLM’s technical capabilities but also by the quality of the human-AI therapeutic relationship, the engagement of content, and the clarity of communication~\cite{li2023systematic,wang2025evaluating}. LLMs have also been applied for measurable goal generation in treatment planning for severe mental illness, showing the potential for integration into structured care workflows~\cite{james2023towards}. Together, these findings suggest that LLMs can serve as valuable clinical aids, augmenting human professionals and supporting scalable mental health interventions, provided that oversight and evaluation frameworks are carefully implemented.

\subsection{Risks and Challenges of LLMs in Mental Health}
Despite the promise of LLMs, numerous risks and challenges persist in their application to mental health. Technical limitations include hallucinations (i.e. providing inaccurate information), limited interpretability, inconsistent predictions, and biases in both generated content and response quality~\cite{guo2024large,chung2023challenges,yuan2025improving,aleem2024towards,ji2023rethinking}. Cultural sensitivity, empathy, and adaptability remain areas of concern, with studies highlighting deficits in engagement depth, listening, and the ability to contextualize user needs~\cite{aleem2024towards,ma2024understanding}. Ethical issues are also particularly concerning, encompassing privacy violations, lack of informed consent, unclear accountability, and inappropriate trust formation~\cite{cabrera2023ethical,lawrence2024opportunities,kwesi2025exploring,guo2024large}. Overreliance on LLMs can lead to emotional dependency, misinterpretation of user disclosures, and potential erosion of therapeutic rapport~\cite{hua2024large,kwesi2025exploring,haque2023overview}. Safety concerns also arise regarding LLM-generated advice, power imbalances, and the risk of delivering incorrect guidance~\cite{wang2025evaluating,levkovich2025evaluating,li2025counselbench}. Moreover, LLMs can inadvertently propagate disparities, reinforce stigma, and compromise equity in mental healthcare delivery~\cite{lawrence2024opportunities,gabriel2024can}. Researchers have emphasized that, despite the capabilities of LLMs, human counselors’ nuanced understanding, empathy, and contextual judgment remain irreplaceable~\cite{ji2023rethinking}, and that rigorous ethical guidelines, professional oversight, and robust evaluation frameworks are essential for safe deployment~\cite{cabrera2023ethical,guo2024large,lawrence2024opportunities}.

\subsection{Summary}
Despite this growing body of work that explores various applications of LLMs and digital conversational agents in mental health, there remains a lack of systematic comparative studies evaluating LLM performance against human therapists. Our work addresses this gap by evaluating answers to real patient questions from both perspectives, incorporating ratings from end users as well as professional therapists. This approach allows us to assess not only the perceived quality and helpfulness of LLM outputs, but also how they align with clinical standards, providing a clearer understanding of the strengths and limitations of LLMs in supporting mental health care.

\section{Method}
To address our research questions, we designed a survey to collect user ratings of both therapist-written and LLM-generated answers to mental health-related questions. Our institution's Institutional Review Board (IRB) approved this study, and all participants provided informed consent prior to their participation. 

\subsection{Recruitment and Participants}
To avoid potential bias, such as participants' distrust of LLMs, we framed the task as ``rating different answers to mental health-related questions'' and did not mention the potential use of LLMs. We recruited regular users and licensed therapists through different procedures to accommodate their availability and professional background. We recruited regular users via Prolific~\footnote{https://prolific.com}, a crowd-sourcing platform commonly employed for research studies. We conducted recruitment in November 2024. The survey required approximately 15 minutes to complete, and participants received \$10 in compensation. We included responses from a total of 150 participants in the final dataset and present their demographic information in Table~\ref{tab:user_demographics}. 

We also recruited licensed therapists between November 2024 and January 2025 through internal listservs and snowball sampling. Public platforms do not suit recruiting licensed therapists, as they lack reliable means to verify participants' professional credentials. Finally, a total of 23 licensed therapists completed the survey, and we present their demographic information in Table~\ref{tab:therapist_demographics}. As compensation, we entered therapist participants into a lottery, and a selected individual received a \$100 Amazon gift card.

\begin{table*}[htbp]
\centering
\begin{minipage}{0.48\textwidth}
\centering
\begin{tabular}{lll}
\toprule
\textbf{Category} & \textbf{Variable} & \textbf{Percentage} \\
\hline
\multirow{5}{*}{\textbf{Age}} & 18-24 years old & 20.7 \\
 & 25-34 years old & 39.3 \\
 & 35-44 years old & 21.3 \\
 & 45-54 years old & 13.3 \\
 & 55 years old or older & 4.7 \\
\hline
\multirow{3}{*}{\textbf{Gender}} & Male & 49.3 \\
 & Female & 48.7 \\
 & Non-binary / third gender & 2.0 \\
\hline
\multirow{6}{*}{\textbf{Education}} & High school graduate & 10.7 \\
 & Some college & 22.7 \\
 & 2 year degree & 7.3 \\
 & 4 year degree & 36.7 \\
 & Professional degree & 20.7 \\
 & Doctorate & 2.0 \\
\hline
\multirow{5}{*}{\textbf{Experience}} 
 & No experience & 29.3 \\
 & Under a year & 22.7 \\
 & 5 - 10 years & 10.0 \\
 & Over 10 years & 6.0 \\
 & 1 - 5 years & 32.0 \\
\bottomrule
\end{tabular}
\caption{User participant demographics information.}
\label{tab:user_demographics}
\end{minipage} \hfill \begin{minipage}{0.48\textwidth}
\centering
\begin{tabular}{lll}
\toprule
\textbf{Category} & \textbf{Option} & \textbf{Percentage} \\
\hline
\multirow{4}{*}{\textbf{Age}} & 25-34 years old & 39.1 \\
 & 35-44 years old & 26.1 \\
 & 45-54 years old & 13.0 \\
 & 55 years old or older & 21.7 \\
\hline
\multirow{2}{*}{\textbf{Gender}} & Male & 34.8 \\
 & Female & 65.2 \\
\hline
\multirow{2}{*}{\textbf{Education}} & Professional degree & 43.5 \\
 & Doctorate & 56.5 \\
\hline
\multirow{5}{*}{\textbf{Experience}} & Under a year & 4.3 \\
 & 1 - 2 years & 8.7 \\
 & 2 - 5 years & 17.4 \\
 & 5 - 10 years & 21.7 \\
 & Over 10 years & 47.8 \\
\bottomrule
\end{tabular}
\caption{Therapist participant demographics information.}
\label{tab:therapist_demographics}
\end{minipage}
\end{table*}

\subsection{Survey Design}
\subsubsection{Dataset}
To approximate real-world use, we used Counsel Chat, a dataset of mental health–related questions posted by users and answered by licensed therapists~\cite{counselchat}. This anonymized dataset was collected through an online platform where users could ask mental health questions and expect responses from licensed mental health professionals in a publicly visible, message board-type format. 
After filtering out questions without therapist responses, we retained 845 questions, many of which had multiple counselor answers and collectively covered more than 30 different topics. Those topics included but were not limited to, depression, anxiety, and relationship issues. From this dataset, we extracted the unique questions and used them as prompts in conversations with LLM chatbots to collect LLM-generated responses. We deliberately avoided using system-level prompting or additional instructions, since our goal was to replicate the conditions under which everyday users would engage with these tools, rather than optimizing responses for research purposes.

To capture the variety of models available to the public, we selected three widely used LLMs: ChatGPT, Gemini, and Llama (see Appendix~\ref{sec:model_specification} for model specification). This combination allowed us to examine both proprietary, commercially deployed models (ChatGPT, Gemini) and open-source alternatives (Llama). Because chatbot responses can differ across sessions, we prompted each model three separate times for every question, each time starting a new session. This approach accounts for variability in LLM behavior while maintaining consistency across models, yielding three responses per question per model.

\subsubsection{Questionnaire}
To ensure a manageable study size, we randomly sampled 90 questions from the dataset while keeping the proportional distribution of topics consistent with the full corpus. For each question, we randomly assigned one of the three LLM models and randomly selected one response from the three pre-generated outputs. Each model was therefore represented with 30 unique questions. When multiple therapist answers were available, we selected one at random to pair with the LLM-generated response. This process yielded 90 questions, each with one therapist-written and one LLM-generated answer. All responses were manually examined to ensure that no harmful content was included. In the survey, participants were asked to evaluate three questions, each paired with responses from different LLMs, with questions assigned randomly to ensure balanced representation across models.
We presented each question twice in a row within the survey, once with a therapist response and once with an LLM-generated response, in randomized order. For each response, participants rated it along the following dimensions on a five-point Likert scale ranging from strongly disagree to strongly agree: 
\begin{itemize}
    \item This response is clear. 
    \item This response answers the question. 
    \item This response is encouraging and supportive. 
    \item This response is respectful, accepting, and not judging. 
    \item Overall, I like this answer.
\end{itemize}

These evaluation dimensions reflect qualities central to effective communication in mental health contexts. We derived them from common, nonspecific factors known to contribute to therapeutic alliance and perceived helpfulness~\cite{nonspecific2011, Wampold2015Great}. While our study focuses on answering mental health–related questions rather than conducting therapy, such answers cannot be judged solely on factual correctness; they also require qualities similar to those valued in effective therapeutic interactions. To properly refine and adapt these dimensions for our study, two experimental psychologists on our research team (one of whom is a licensed mental health professional) reviewed and discussed their appropriateness. The procedure for licensed therapist participants was identical, except that they also rated each response on an additional dimension of whether the response was acceptable from a professional perspective.  

For each participant, the same question-response pairs were then presented again, and they were asked to indicate how likely they believed each response had been generated by an LLM. This task was given only after all rating tasks were completed to minimize potential bias. The full survey can be found in Appendix~\ref{sec:survey}.

Because survey responses could be lengthy and cognitively demanding to read, we included attention check questions to ensure data quality. Each participant received one attention check question, which was randomly embedded into the survey. The attention check appeared as an additional evaluation dimension and asked whether the current question was on a specific topic, with the topic label drawn from the Counsel Chat dataset. Participants were expected to select ``Agree'' or ``Strongly agree'' to pass. Only two user participants failed the attention check questions, and we excluded their responses from the final dataset. 

\subsection{Data Analysis}
To establish a comprehensive comparison between human therapist answers and LLM-generated answers, we analyzed a variety of textual properties as well as participant ratings on those answers.

\subsubsection{Text Analysis}
To characterize differences between therapist and LLM language use, we examined several textual properties that capture length, style, and tone. Response length was measured by word count, while readability was assessed using the Flesch Reading Ease score, a standard index of linguistic accessibility~\cite{flesch_new_1948}. To measure vocabulary richness, we calculated a normalized type-token ratio, which divides the number of unique words by the square root of the total words in a response in order to reduce length bias. We also quantified stylistic differences by identifying hedging expressions such as ``possibly'' or ``it depends'', which signal tentativeness, and by counting first-person pronouns, which reflect subjective framing. Finally, we quantified the sentiment of the responses using the VADER sentiment lexicon, which provides a compound sentiment score ranging from negative to positive~\cite{vadar}.
For each of these measures, we compared distributions across therapist and LLM responses. We evaluated statistical significance using Mann–Whitney U tests~\cite{mann1947test} and rank-biserial correlation~\cite{kerby2014simple} for effect size, allowing us to identify whether observed differences are significant without assuming normal distributions.

\subsubsection{Survey Response Analysis}
To compare ratings of therapist and LLM answers, we constructed three therapist–LLM pairs, corresponding to the three LLM models under study. We applied the Wilcoxon signed-rank test, which is a non-parametric alternative to the paired t-test, appropriate for pairwise comparisons of ordinal data that do not meet normality assumptions. Specifically, for each evaluation dimension, we tested whether the paired difference between LLM and therapist scores was greater than zero (one-sided test) with $D^{\text{Dimension}}_{\text{Model}} = S^{\text{Dimension}}_{\text{Model}} - S^{\text{Dimension}}_{\text{Human}}$.



To investigate how participant ratings were affected by different factors, we modeled the difference in ratings between LLM-generated answers and therapist-written answers. This difference served as our dependent variable. To analyze these ordered differences while accounting for repeated measurements from the same participants, we used a cumulative link mixed model (CLMM). A CLMM is an extension of ordinal regression that handles ordered categorical outcomes and incorporates random effects to model within-participant correlation. Intuitively, the model estimates the probability that a rating difference falls at or below each cutoff on the scale, such as whether a participant rated the therapist-written response higher than the LLM-generated one, rated them equally, or rated the LLM-generated answer higher.

The primary predictor of interest was the model used for the response (i.e., ChatGPT, Gemini, or Llama), which we refer to as treatment later in the paper. This allowed us to directly compare participant ratings across outputs from different LLMs. The model also included demographic covariates (education, age, race, state, gender) and experience undergoing therapy, as well as interaction terms to test whether the effect of treatment varied by experience undergoing therapy. Participant-specific random intercepts were included to account for repeated ratings from the same individuals.

Additionally, we examined participants’ judgments of answer authorship. For each response, participants were asked how much they agreed with the statement that the answer was generated by an AI. Selecting ``Strongly agree'' or ``Somewhat agree'' was coded as agreement that the response was LLM-generated, whereas ``Strongly disagree'' or ``Somewhat disagree'' was coded as agreement that the response was therapist-written. A ``Neutral'' selection was treated as confusion or a lack of confident judgment. Whether participants can distinguish therapist-written and LLM-generated answers correctly was also included as a covariate of the CLMM to explore whether participants’ ability to detect AI-generated content moderated their ratings.

All statistical analyses were conducted in R, using the \verb|base| stats package for the Wilcoxon signed-rank test and the \verb|ordinal| package for CLMM. Full mathematical details of the CLMM specification are provided in the Appendix~\ref{sec:statistics}.

\subsection{Limitations}
Our study has several limitations. First and most importantly, it was not a direct test of therapist responses to their own patients relative to chatbot answers; instead, we compared how licensed therapists and LLMs responded to mental health–related questions in the context of a public forum (i.e. Counsel Chat), where the quality and depth of these responses may vary and may not fully reflect professional therapeutic standards. As such, the findings have limited applicability to therapy provision, where the therapeutic relationship and context differ significantly. Our study is best understood as examining the potential of chatbots to provide adjunctive, informational support for mental health rather than substitutes for professional care. Moreover, participant ratings reflect perceptions of qualities such as clarity, empathy, and respect rather than verified therapeutic efficacy. Responses rated poorly may still hold superior therapeutic value, and understanding how these crowd-sourced ratings relate to actual therapeutic efficacy remains an open research question.

Second, our study design and sample introduce constraints on generalizability. All participants were English-speaking U.S. adults, which may limit applicability to other cultures, languages, and demographic groups. Our therapist sample was also relatively small (n = 23), reducing statistical power for comparisons between professional and non-professional evaluations. In addition, we deliberately avoided prompt engineering or system-level customization for LLM-generated answers to approximate everyday usage, but this may underestimate system performance under optimized conditions. Finally, our evaluation focused on short-term perceptual ratings, which capture important aspects of supportive communication but do not address therapeutic safety, long-term effectiveness, or risks.

\section{Findings}
In this section, we first present our results for text analysis, followed by those for the survey responses. 

\subsection{Text Analysis}
We compared therapist and LLM responses to mental health-related questions across multiple textual and stylistic metrics, including length, readability, type-token ratio, hedging, first-person usage, and sentiment. We assessed statistical significance using both independent samples t-tests and Mann–Whitney U tests, with results summarized in Table~\ref{table:p_values}.

\paragraph{Answer Length and Complexity.}
\begin{figure*}[ht]
\noindent
\centering
\begin{minipage}{.34\textwidth}
\centering
    \includegraphics[width=\linewidth]{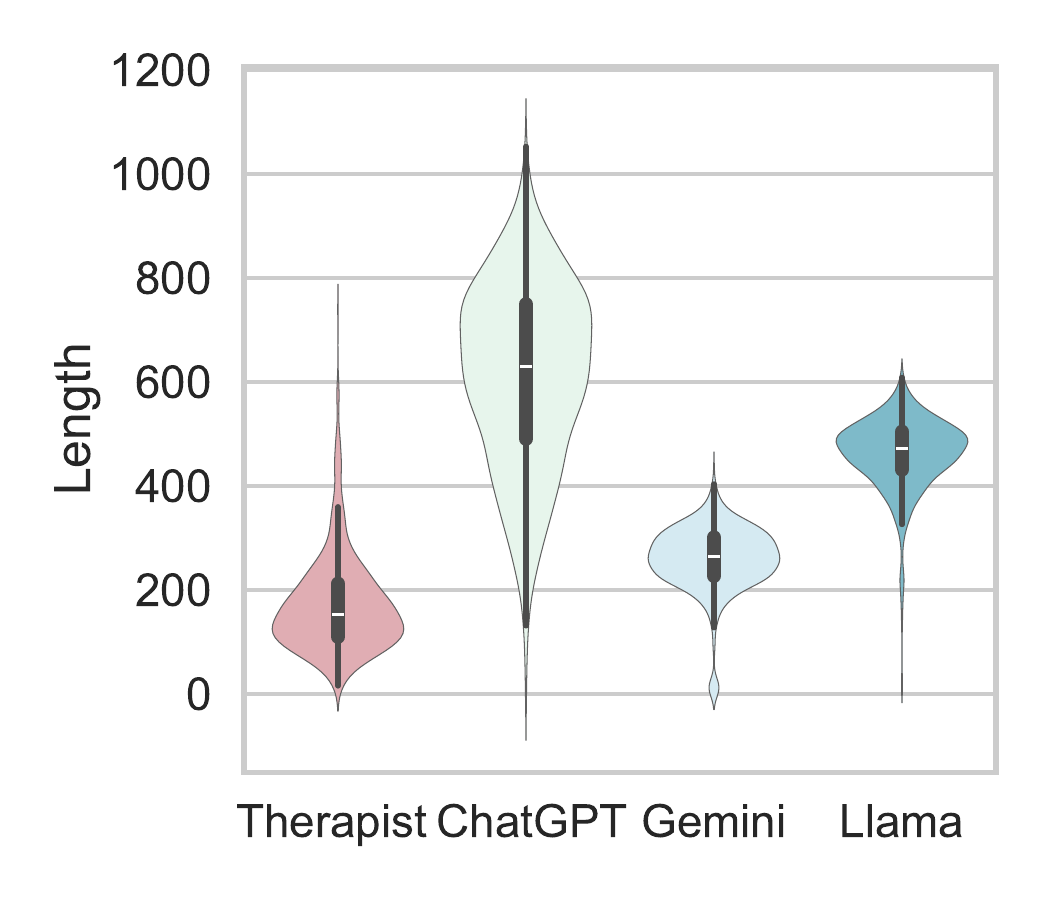}
    \caption{Distribution of length.}
    \label{fig:length}
\end{minipage}\begin{minipage}{.34\textwidth}
\centering
    \includegraphics[width=\linewidth]{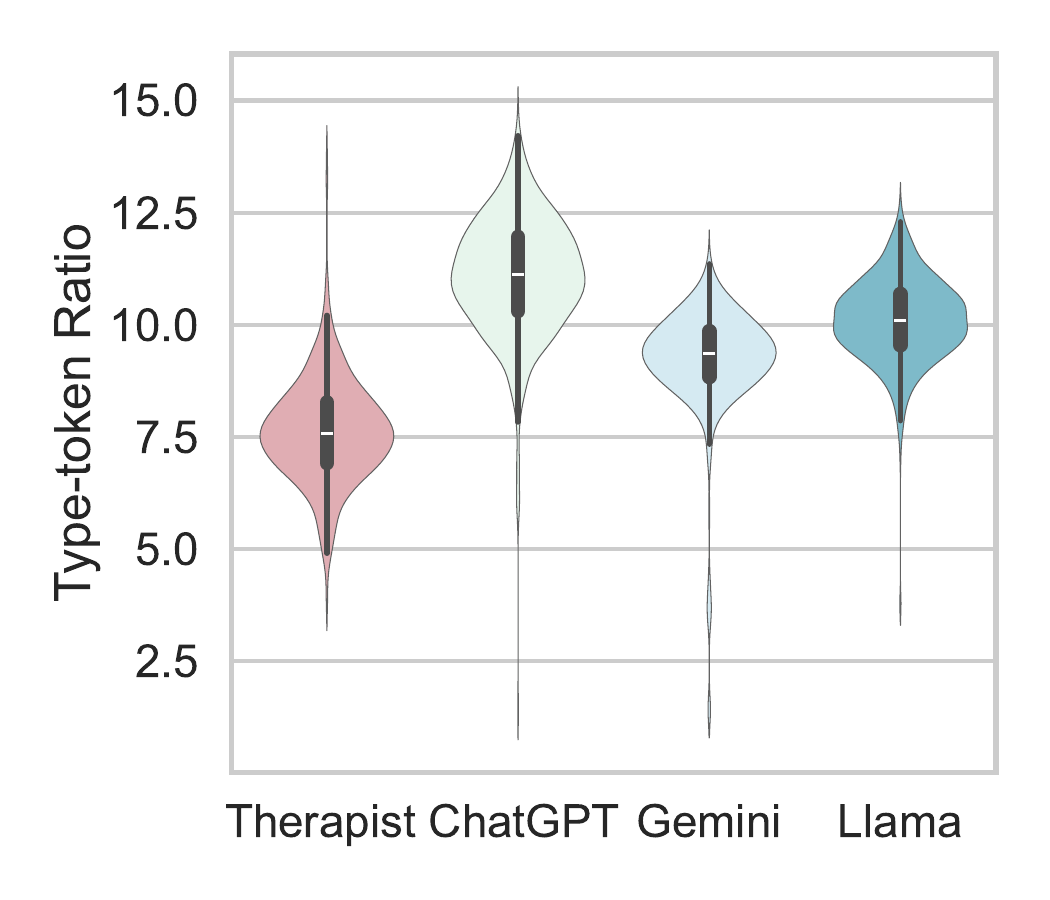}
    \caption{Distribution of type-token ration.}
    \label{fig:ttr}
\end{minipage}\begin{minipage}{.34\textwidth}
\centering
    \includegraphics[width=\linewidth]{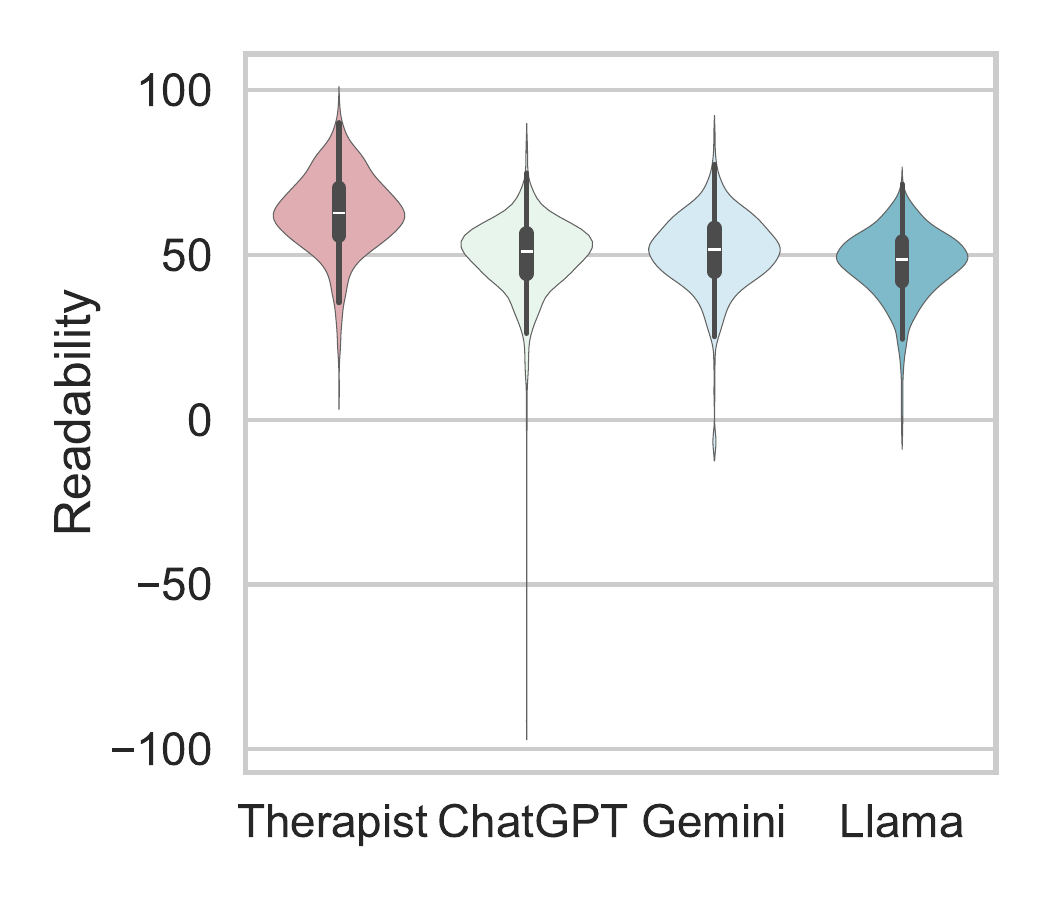}
    \caption{Distribution of readability score.}
    \label{fig:readability}
\end{minipage}
\end{figure*} 
As illustrated in Figure~\ref{fig:length}, ChatGPT answers are the longest with the highest variation in length. All three LLMs produced responses that were significantly longer than therapist answers ($p < 0.001$) for Mann–Whitney U tests. Similarly, LLM answers have significantly higher type-token ratios, which indicate higher vocabulary diversities (Figure~\ref{fig:ttr}), suggesting that LLMs tend to produce more elaborated and lexically varied text compared to therapists. However, therapist responses were significantly more readable, as demonstrated by their higher readability score in Figure~\ref{fig:readability}.

\paragraph{Hedging and Subjectivity.}
\begin{figure*}[ht]
\noindent
\centering
\begin{minipage}{.34\textwidth}
\centering
    \includegraphics[width=\linewidth]{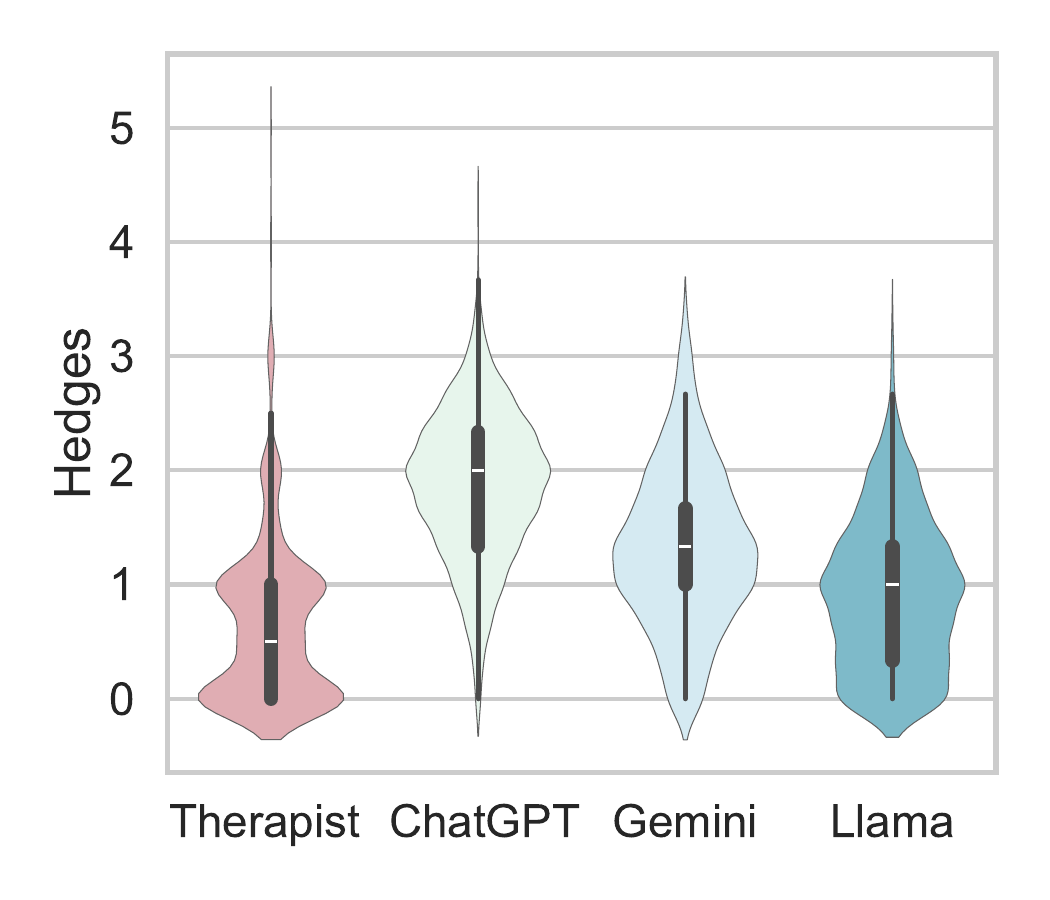}
    \caption{Distribution of hedge word count.}
    \label{fig:hedge}
\end{minipage}\begin{minipage}{.34\textwidth}
\centering
    \includegraphics[width=\linewidth]{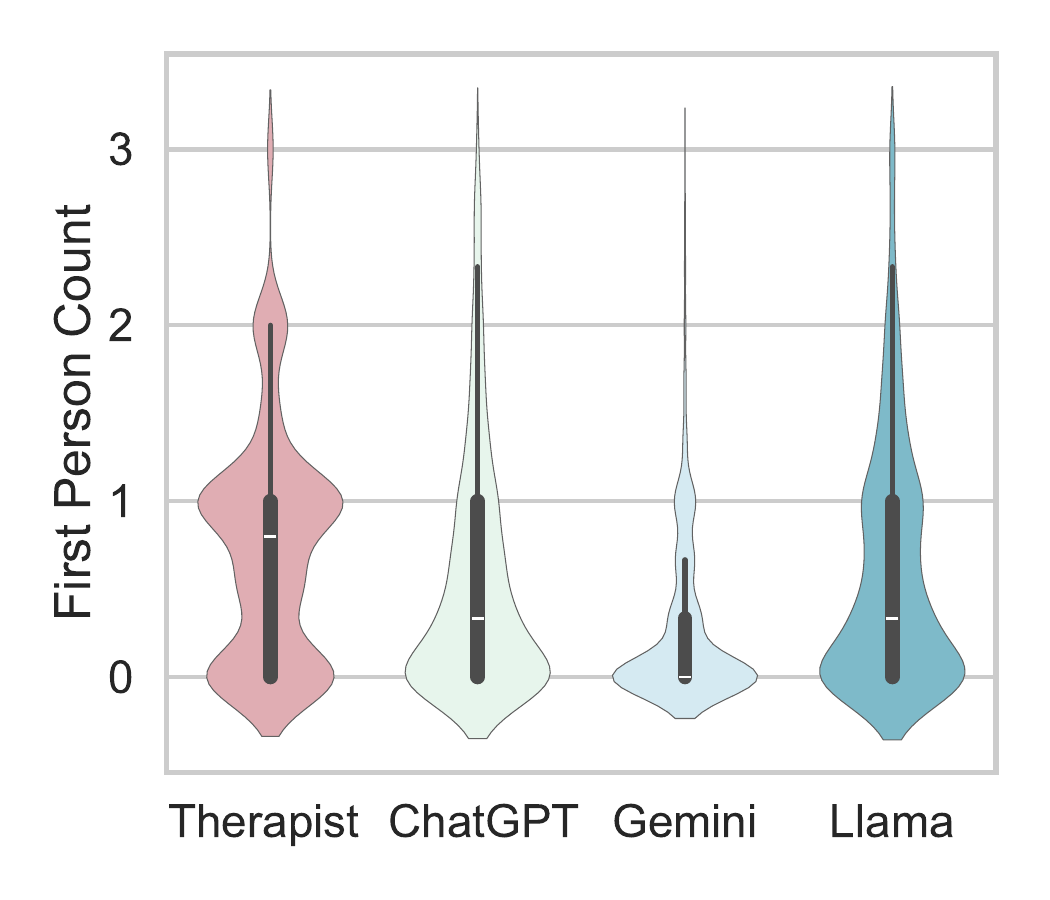}
    \caption{Distribution first person word count.}
    \label{fig:subjectivity}
\end{minipage}\begin{minipage}{.34\textwidth}
\centering
    \includegraphics[width=\linewidth]{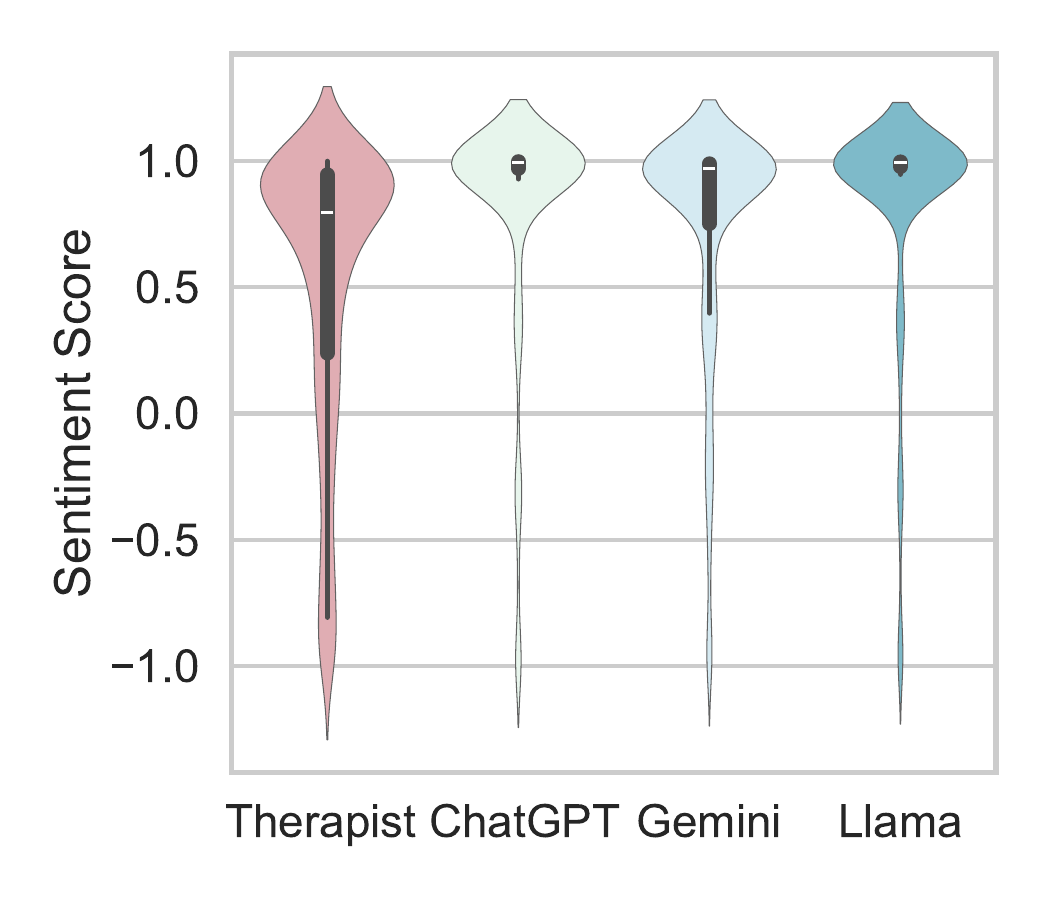}
    \caption{Distribution of sentiment score.}
    \label{fig:sentiment}
\end{minipage}
\end{figure*}
LLM responses contained significantly more hedge words than therapist answers across all models (Figure~\ref{fig:hedge}), indicating a tendency toward cautious or qualified language. In contrast, therapists used first-person references more frequently (Figure~\ref{fig:subjectivity}), reflecting a more subjective and personal style of response. We found these differences to be statistically significant ($p < 0.001$) for all models.

\paragraph{Sentiment.}
LLM-generated responses exhibited slightly higher sentiment scores than therapist responses (Figure~\ref{fig:sentiment}), with significant differences between distributions ($p < 0.001$), suggesting a more positive or neutral tone overall.

In summary, LLMs generated responses that are longer and lexically richer, with more hedging and a more neutral tone, while therapists produced shorter, more readable responses with more frequent usage of first-person framing. We observed these differences consistently across all three LLM models and found them to be statistically significant.

\newcolumntype{d}[1]{D{.}{.}{#1}}

\begin{table*}[!ht]
\begin{tabular}{llS @{\hskip 3em} d{2}}
\toprule
Metric & Model & $p_u$ & \multicolumn{1}{c}{Effect Size} \\
\midrule
\multirow{3}{*}{Length} & ChatGPT & 1.82e-267*** & 0.96 \\
 & Gemini & 1.88e-126*** & 0.66 \\
 & Llama & 3.85e-262*** & 0.95 \\
\midrule
\multirow{3}{*}{Readability} & ChatGPT & 4.34e-114*** & -0.62 \\
 & Gemini & 7.23e-95*** & -0.57 \\
 & Llama & 8.37e-149*** & -0.71 \\
\midrule
\multirow{3}{*}{Type-token Ratio} & ChatGPT & 4.33e-259*** & 0.94 \\
 & Gemini & 6.27e-175*** & 0.77 \\
 & Llama & 2.45e-243*** & 0.91 \\
\midrule
\multirow{3}{*}{Hedges} & ChatGPT & 2.91e-188*** & 0.79 \\
 & Gemini & 1.04e-87*** & 0.54 \\
 & Llama & 5.59e-23*** & 0.27 \\
\midrule
\multirow{3}{*}{First Person Count} & ChatGPT & 8.79e-15*** & -0.21 \\
 & Gemini & 5.83e-65*** & -0.44 \\
 & Llama & 8.81e-08*** & -0.14 \\
\midrule
\multirow{3}{*}{Sentiment Score} & ChatGPT & 2.24e-123*** & 0.65 \\
 & Gemini & 4.08e-51*** & 0.41 \\
 & Llama & 6.84e-134*** & 0.67 \\
\bottomrule
\end{tabular}
\setlength{\abovecaptionskip}{8pt plus 3pt minus 2pt}
\caption{Mann–Whitney U test and effect size results.}
\vspace{-5pt}
\label{table:p_values}
\end{table*}

\subsection{User Ratings}
\begin{figure*}[ht]
\noindent
\centering
\begin{minipage}{.32\textwidth}
\centering
    \includegraphics[width=\linewidth]{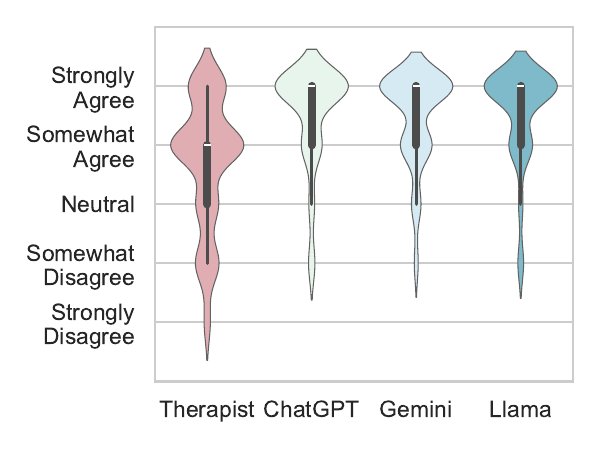}
    \setlength{\abovecaptionskip}{-8pt plus 3pt minus 2pt}
    \caption{User ratings for ``this response is clear.''}
    \label{fig:user_1}
\end{minipage}\hfill\begin{minipage}{.32\textwidth}
\centering
    \includegraphics[width=\linewidth]{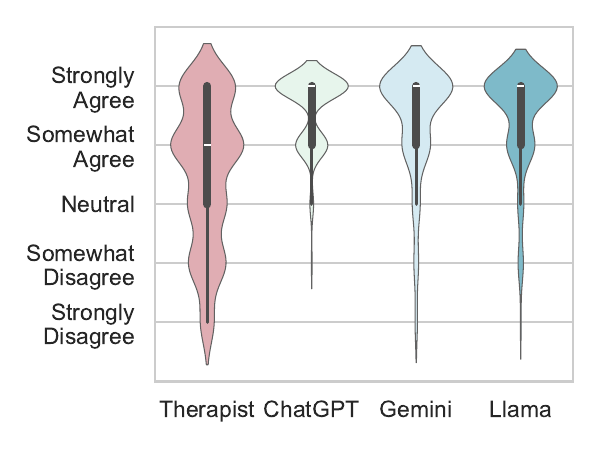}
    \setlength{\abovecaptionskip}{-8pt plus 3pt minus 2pt}
    \caption{User ratings for ``this response answers the question.''}
    \label{fig:user_2}
\end{minipage}\hfill\begin{minipage}{.32\textwidth}
\centering
    \includegraphics[width=\linewidth]{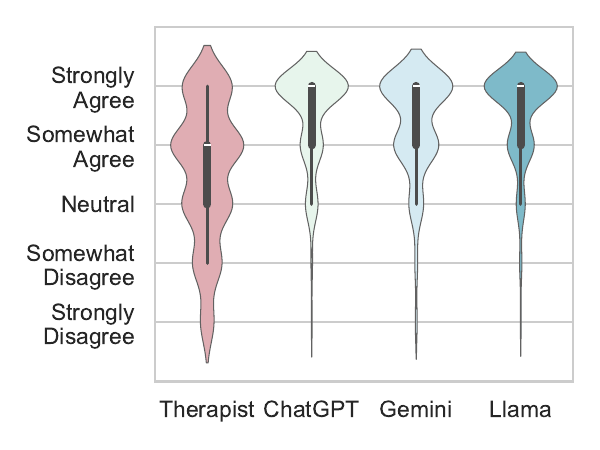}
    \setlength{\abovecaptionskip}{-8pt plus 3pt minus 2pt}
    \caption{User ratings for ``this response is encouraging and supportive.''}
    \label{fig:user_3}
\end{minipage}
\centering
\begin{minipage}{.32\textwidth}
\centering
    \includegraphics[width=\linewidth]{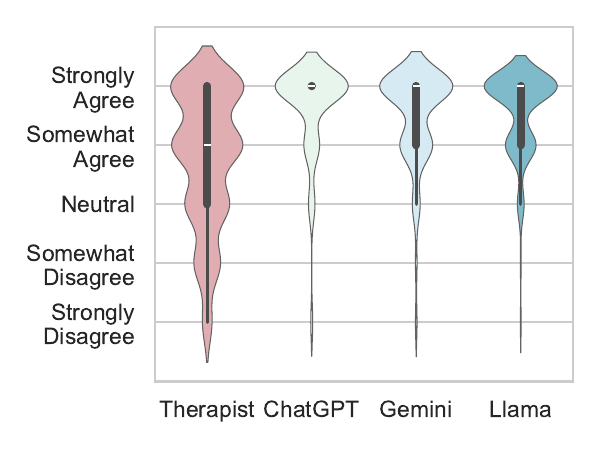}
    \setlength{\abovecaptionskip}{-8pt plus 3pt minus 2pt}
    \caption{User ratings for ``this response is respectful, accepting, and not judging.''}
    \label{fig:user_4}
\end{minipage}\hspace{10pt}\begin{minipage}{.32\textwidth}
\centering
    \includegraphics[width=\linewidth]{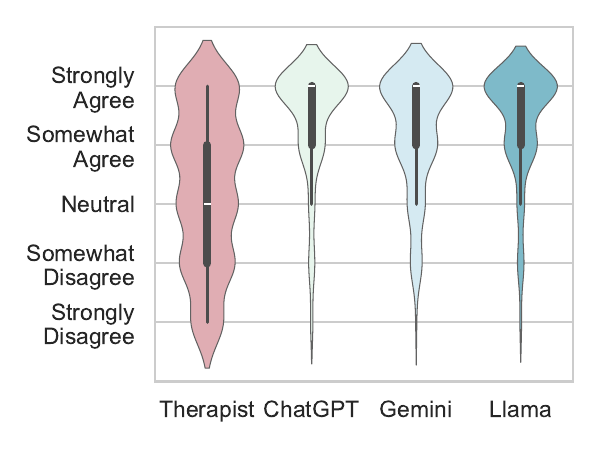}
    \setlength{\abovecaptionskip}{-8pt plus 3pt minus 2pt}
    \caption{User ratings for ``overall, I like this answer.''}
    \label{fig:user_5}
\end{minipage}
\end{figure*}

Distributions of user ratings for each dimension are shown in Figure~\ref{fig:user_1} -~\ref{fig:user_5}.
Across all dimensions, participants consistently rated LLM-generated answers higher than those of human therapists. As reported in Table~\ref{table:user_rating_p}, these differences are statistically significant: all one-sided p-values are below $10^{-7}$, remaining robust even under a conservative Bonferroni correction with an adjusted threshold of $p < 0.001$ to account for multiple testing. 

\begin{table*}[h!]
\centering
\begin{tabular}{clS}
\toprule
\textbf{Model} & \textbf{Variable} & \textbf{P Value} \\
\midrule
ChatGPT & This response is clear.                                      & 8.20e-08*** \\
        & This response answers the question.                          & 1.16e-13*** \\
        & This response is encouraging and supportive.                 & 1.14e-10*** \\
        & This response is respectful, accepting, and not judging.     & 9.87e-09*** \\
        & Overall, I like this answer.                                 & 6.16e-10*** \\
\midrule
Gemini  & This response is clear.                                      & 1.15e-11*** \\
        & This response answers the question.                          & 4.18e-10*** \\
        & This response is encouraging and supportive.                 & 2.43e-09*** \\
        & This response is respectful, accepting, and not judging.     & 1.14e-08*** \\
        & Overall, I like this answer.                                 & 7.00e-09*** \\
\midrule
Llama   & This response is clear.                                      & 2.74e-12*** \\
        & This response answers the question.                          & 4.15e-12*** \\
        & This response is encouraging and supportive.                 & 3.68e-15*** \\
        & This response is respectful, accepting, and not judging.     & 1.72e-15*** \\
        & Overall, I like this answer.                                 & 8.42e-14*** \\
\bottomrule
\end{tabular}
\vspace{8pt}
\caption{This table presents the one-sided p-values of Wilcoxon signed-rank tests testing whether the rating by participants for each variable in the LLM model response is significantly higher than that of the corresponding human response.}
\label{table:user_rating_p}
\end{table*}

\subsubsection{Model Comparison}
Results from the CLMM (Table~\ref{tab:clmm}) indicate that, after controlling for participants' demographic characteristics and their ability to judge authorship (i.e., knowing whether the response was from a human or LLM), Gemini responses were rated significantly lower than those of ChatGPT ($\alpha_{\text{gemini}} = -1.16, p=0.002$). This corresponds to an odds ratio of approximately 0.31, suggesting that participants were substantially likely to rate Gemini outputs lower. In contrast, we failed to find a difference between Llama and ChatGPT, which suggests comparable overall performance.



\begin{table*}[htbp]
\centering
\begin{tabular}{ll @{\hskip 3em}l  @{\hskip 3em}r @{\hskip 3em}r}
\toprule
Category & Predictor & \text{Estimate} & \text{Std. Error} & \text{$p$-value} \\
\midrule
\multirow{2}{*}{Treatment}&  Gemini & $-1.164^{**}$ & 0.380 & 0.002 \\
& Llama & $-0.399$ & 0.380 & 0.294 \\  \midrule
\multirow{4}{*}{Therapy experience} & $<$1 year & $-1.521^{**}$ & 0.474 & 0.001 \\
& 1--5 years & $-0.826^{.}$ & 0.425 & 0.052 \\
& 5--10 years & $-1.085^{.}$ & 0.626 & 0.083 \\
& $>$10 years & $-1.107$ & 0.763 & 0.147 \\  \midrule
\multirow{2}{*}{Interaction effect} & Gemini $\times$ $<$1 year & $+1.753^{**}$ & 0.581 & 0.002 \\
& Llama $\times$ $>$10 years & $+2.561^{**}$ & 0.933 & 0.006 \\  \midrule
Authorship judgment & Can distinguish & $-0.224$ & 0.202 & 0.267 \\
\midrule
Random intercept variance (ID) & \multicolumn{4}{c}{0.283} \\
\bottomrule
\end{tabular}
\vspace{8pt}
\caption{Cumulative Link Mixed Model results. Baseline categories: ChatGPT (Treatment), 
``Never received services'' (Therapy Experience).
Significance codes: $^{***} p<0.001$, $^{**} p<0.01$, $^{*} p<0.05$, $^{.} p<0.1$.}
\label{tab:clmm}
\end{table*}


\subsubsection{Demographic Influence}
Participant demographics also influenced ratings, with experience undergoing therapy as a significant predictor. Participants with less than one year of experience consistently rated LLM-generated answers lower ($\beta = -1.52, p=0.001$), while intermediate levels of experience (1–10 years) showed trends toward lower ratings that were marginally significant.


Other demographic factors, including education, age, gender, and participants’ ability to distinguish therapist-written answers from LLM-generated answers (can distinguish), did not significantly predict ratings. These findings suggest that therapist experience is the most pronounced predictor of participant evaluations, while geographic and other demographic effects are less clear, given the limitations of our sample.
\subsubsection{Interaction Effect}
In addition, we observe significant interaction effects between LLM type and participants’ therapist experience on user rating. In other words, experience undergoing therapy changes how participants judge each LLM's responses.  As shown by Table~\ref{tab:clmm}, participants with more than 10 years of therapy experience rated Llama outputs more favorably ($\gamma = +2.56, p=0.006$) compared to ChatGPT outputs rated by participants with no therapist experience, highlighting that the perceived quality of Llama responses increases with professional experience. In contrast, having less than one year of therapy experience has the reversed effect for ratings of Gemini responses ($\gamma = +1.75, p=0.002$), indicating that participants with limited therapy experience rated Gemini more positively than might be expected based on the main effect alone. Overall, these results suggest that while Llama is generally on par with ChatGPT, it is especially well-regarded among participants with long-term therapy experience.

\subsection{Therapist Ratings}
\begin{figure*}[ht]
\noindent
\centering
\begin{minipage}{.32\textwidth}
\centering
    \includegraphics[width=\linewidth]{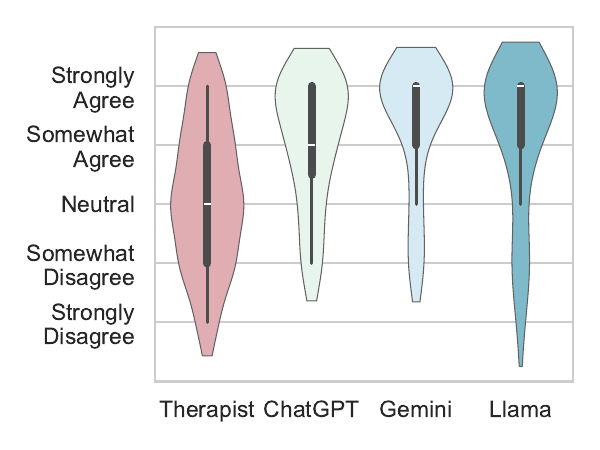}
    \setlength{\abovecaptionskip}{-8pt plus 3pt minus 2pt}
    \caption{Therapist ratings for ``this answer is acceptable from a professional perspective.''}
    \label{fig:therapist_0}
\end{minipage}\hfill\begin{minipage}{.32\textwidth}
\centering
    \includegraphics[width=\linewidth]{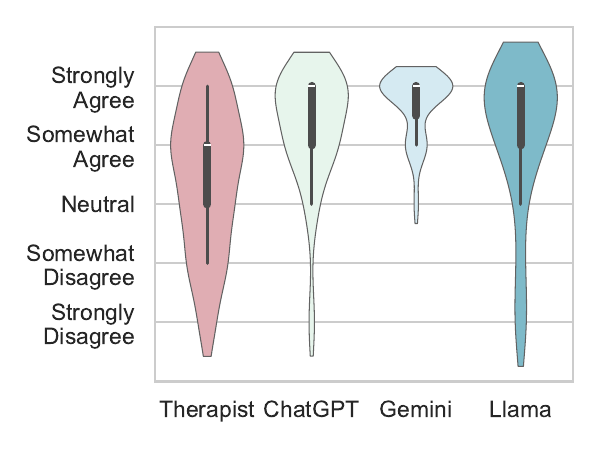}
    \setlength{\abovecaptionskip}{-8pt plus 3pt minus 2pt}
    \caption{Therapist ratings for ``this response answers the question.''}
    \label{fig:therapist_1}
\end{minipage}\hfill\begin{minipage}{.32\textwidth}
\centering
    \includegraphics[width=\linewidth]{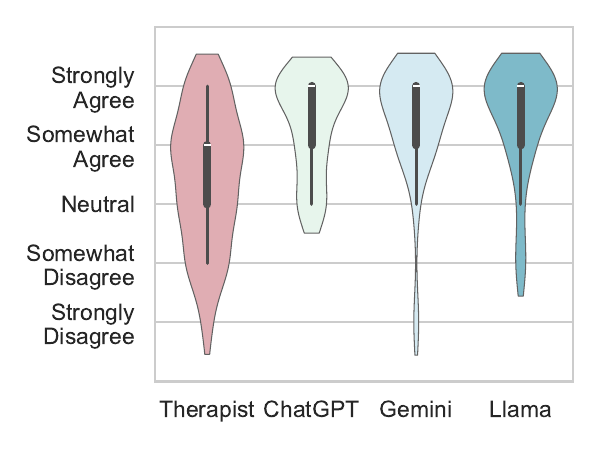}
    \setlength{\abovecaptionskip}{-8pt plus 3pt minus 2pt}
    \caption{Therapist ratings for ``this response is encouraging and supportive.''}
    \label{fig:therapist_2}
\end{minipage}
\centering
\begin{minipage}{.32\textwidth}
\centering
    \includegraphics[width=\linewidth]{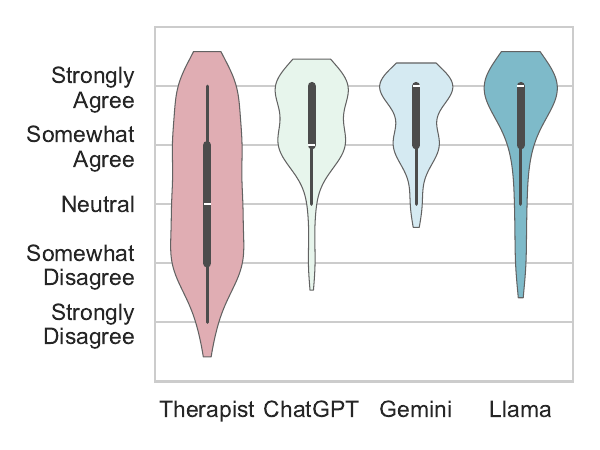}
    \setlength{\abovecaptionskip}{-8pt plus 3pt minus 2pt}
    \caption{Therapist ratings for ``this response is encouraging and supportive.''}
    \label{fig:therapist_3}
\end{minipage}\hfill\begin{minipage}{.32\textwidth}
\centering
    \includegraphics[width=\linewidth]{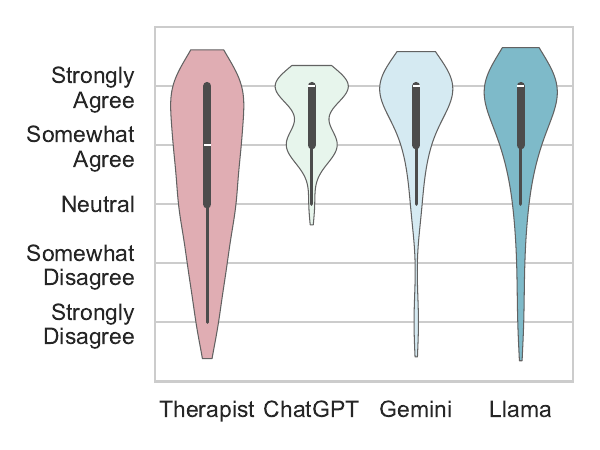}
    \setlength{\abovecaptionskip}{-8pt plus 3pt minus 2pt}
    \caption{Therapist ratings for`` this response is respectful, accepting, and not judging.''}
    \label{fig:therapist_4}
\end{minipage}\hfill\begin{minipage}{.32\textwidth}
\centering
    \includegraphics[width=\linewidth]{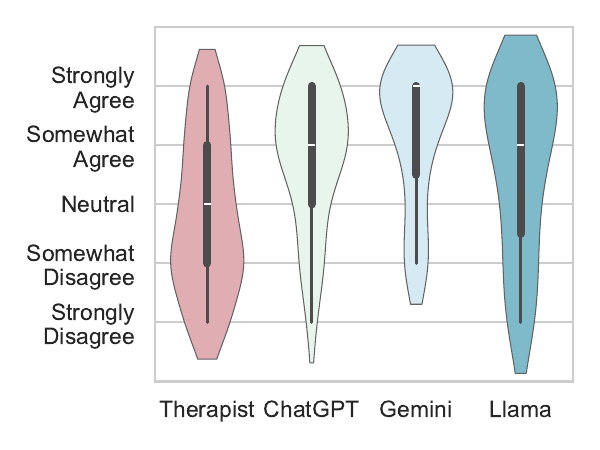}
    \setlength{\abovecaptionskip}{-8pt plus 3pt minus 2pt}
    \caption{Therapist ratings for ``overall, I like this answer.''}
    \label{fig:therapist_5}
\end{minipage}
\end{figure*}

Distributions of therapist ratings for each dimension are shown in Figure~\ref{fig:therapist_0} -~\ref{fig:therapist_5}.
Using the same Wilcoxon signed-rank test we applied to patient ratings, we evaluated therapist ratings across all six dimensions. As shown in Table~\ref{table: therapist_rating_p}, therapists generally rated LLM answers higher, with one-sided p-values indicating statistically significant differences. Therapists rated ChatGPT answers significantly higher than therapist answers across all six dimensions, with $p_1$ values ranging from 0.0156 to 0.00116. Therapists consistently rated Gemini responses significantly higher across all six dimensions, particularly for clarity and professional acceptability ($p_1 < 0.001$). Therapists also rated Llama responses higher for most dimensions, though the differences were smaller and only marginally significant for overall liking ($p_1 = 0.0454$). Although under a conservative Bonferroni correction with an adjusted threshold of $p < 0.001$, many of these results are no longer statistically significant, this correction is likely overly conservative and may obscure meaningful differences, and the general trends remain valid. These findings indicate that, from a professional perspective, LLM-generated responses are generally acceptable and often rated positively, with some variation across models and dimensions, similar to the trends observed in general participant ratings.

\begin{table*}[h!]
\centering
\begin{tabular}{clS}
\toprule
\textbf{Model} & \textbf{Variable} & \textbf{P Value} \\
\midrule
ChatGPT & This response is acceptable from a professional perspective. & 0.015600 * \\
        & This response is clear.                                      & 0.006990 ** \\
        & This response answers the question.                          & 0.001300 ** \\
        & This response is encouraging and supportive.                 & 0.001160 ** \\
        & This response is respectful, accepting, and not judging.     & 0.001860 ** \\
        & Overall, I like this answer.                                 & 0.010200 * \\
\midrule
Gemini  & This response is acceptable from a professional perspective. & 0.000685*** \\
        & This response is clear.                                      & 0.000306*** \\
        & This response answers the question.                          & 0.001750** \\
        & This response is encouraging and supportive.                 & 0.000919*** \\
        & This response is respectful, accepting, and not judging.     & 0.012200* \\
        & Overall, I like this answer.                                 & 0.002830** \\
\midrule
Llama   & This response is acceptable from a professional perspective. & 0.004590** \\
        & This response is clear.                                      & 0.007690** \\
        & This response answers the question.                          & 0.004590** \\
        & This response is encouraging and supportive.                 & 0.001660** \\
        & This response is respectful, accepting, and not judging.     & 0.024700* \\
        & Overall, I like this answer.                                 & 0.045400* \\
\bottomrule
\end{tabular}
\vspace{8pt}
\caption{This table presents the one-sided p-values of Wilcoxon signed-rank tests testing whether the rating by therapists for each variable in the
LLM model response is significantly higher than that of the corresponding human response.}
\label{table: therapist_rating_p}
\end{table*}

We explored the possibility of applying the same CLMM approach to licensed therapist responses in order to compare ratings across treatments and participant demographics. However, we only have 23 therapist responses, which is far too small to reliably estimate the model parameters or include the full set of covariates and random effects. As a result, we did not perform CLMM modeling on this dataset, as any results would be highly unstable and likely misleading.

\subsection{Authorship Judgment}
\begin{figure*}[ht]
\noindent
\begin{minipage}{.49\textwidth}
\centering
    \includegraphics[width=\linewidth]{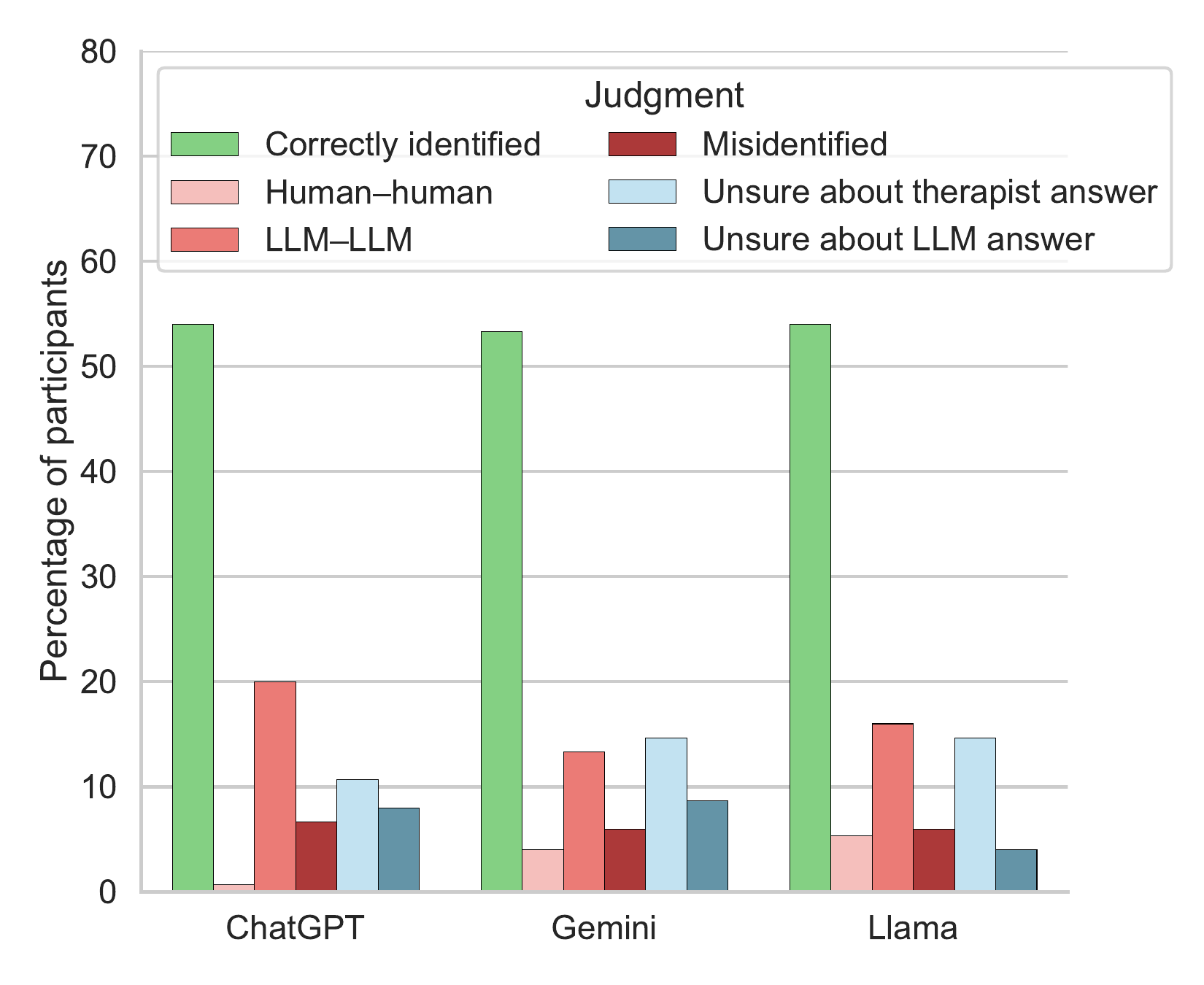}
    \caption{Distribution of user judgment of whether an answer is therapist-written or LLM-generated.}
    \label{fig:user_human_or_not}
\end{minipage}\hfill \begin{minipage}{.49\textwidth}
\centering
    \includegraphics[width=\linewidth]{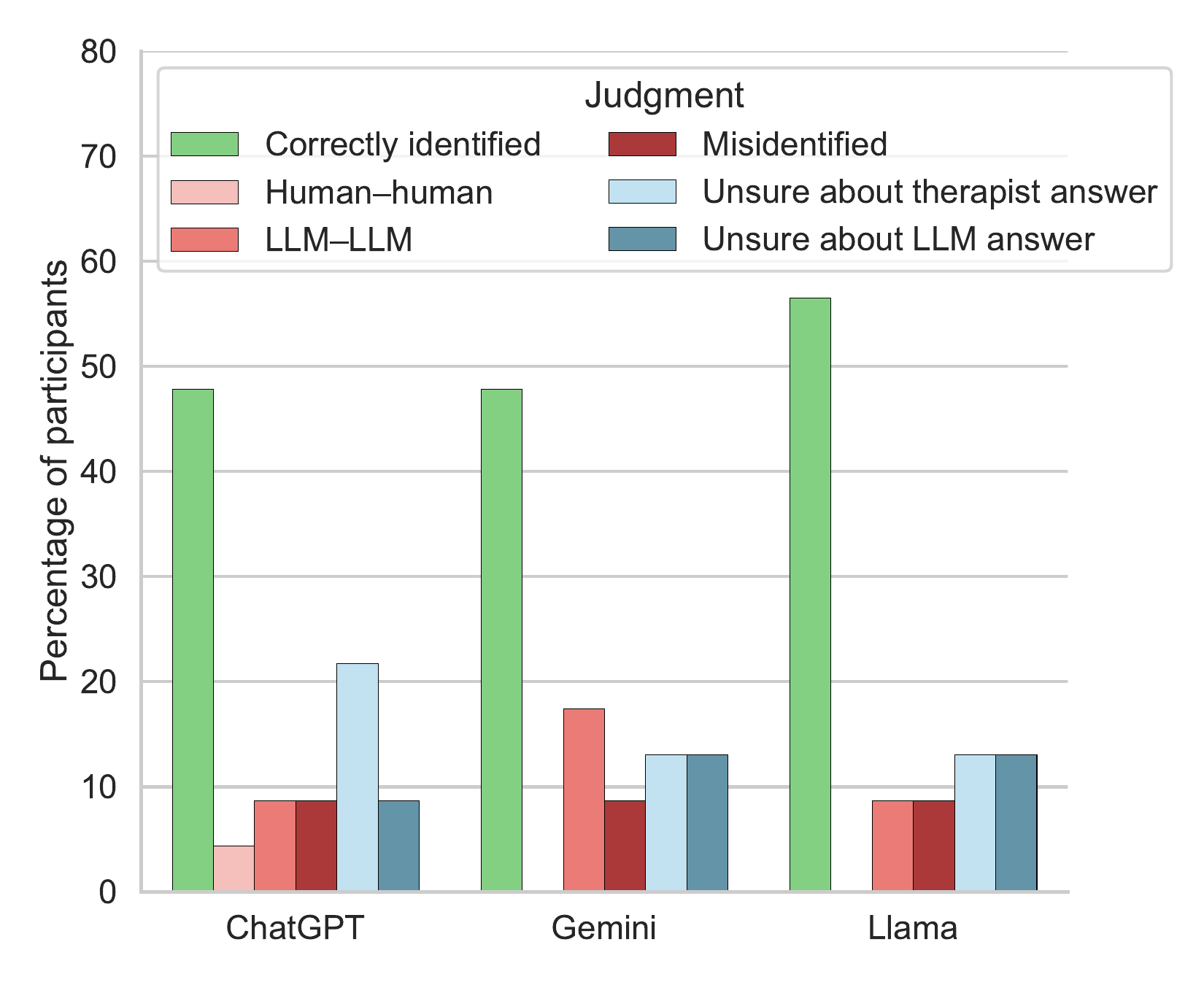}
    \caption{Distribution of therapist judgment of whether an answer is therapist-written or LLM-generated.}
    \label{fig:therapist_human_or_not}
\end{minipage}
\end{figure*}

Across both users and therapists, participants had some difficulty in reliably determining whether an answer was produced by a human therapist or an LLM, but did well above chance. A participant only has a 16\% chance of correctly identifying the authorship of both LLM-generated and therapist-written responses if they were selecting randomly (i.e., selecting two out of the five Likert scale points for both times). As shown in Figure~\ref{fig:user_human_or_not} and~\ref{fig:therapist_human_or_not}, participants correctly identified the authorship of both answers for the same questions about 45\% to 60\% of the time. Some participants attributed both answers to the same source, judging them as either both therapist-written or both LLM-generated. We also observed misclassifications, that is, participants judging therapist answers as LLM-generated and LLM answers as therapist-written. Finally, we found that about 10\% of participants were unsure about the authorship of one or both answers. These patterns suggest that, regardless of background, distinguishing between therapist and LLM responses was, though not straightforward, doable for many participants, reflecting some stylistic similarity between professional and model-generated answers.

In addition, whether participants were able to correctly distinguish authorship did not significantly affect their evaluations of the responses, as our regression analysis showed that the "can distinguish" variable was not a significant predictor of ratings across models, suggesting that participants’ liking and quality judgments were largely independent of their ability to identify whether an answer was written by a therapist or an LLM.

\subsection{User Preferences and Therapist Recommendation}
\begin{figure*}[ht]
\noindent
\centering
    \includegraphics[width=\linewidth]{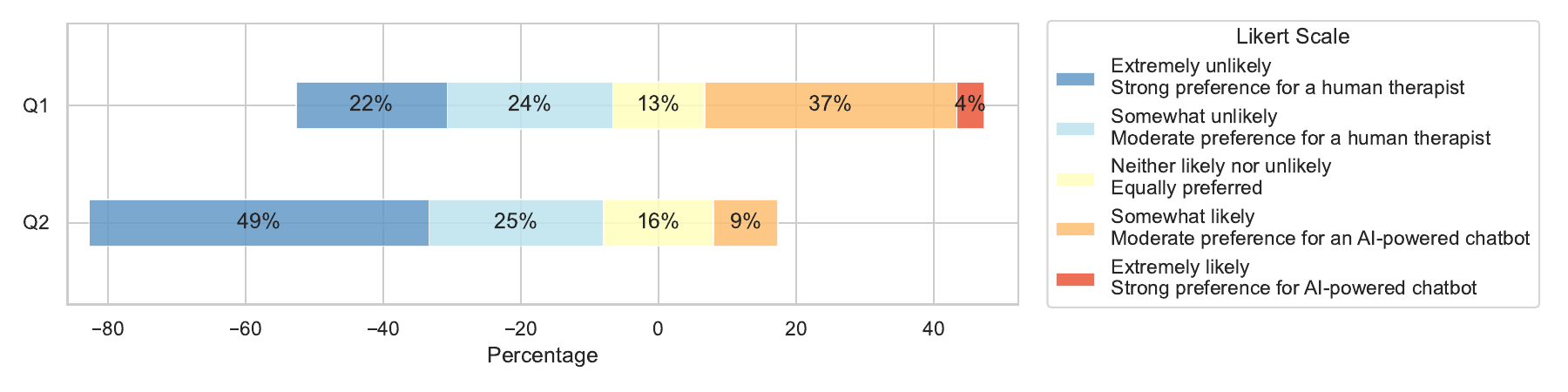}
    \vspace{-15pt}
    \caption{Distribution of user selection for Q1: Would you personally use AI-powered chatbots to answer your mental health-related questions, and Q2: From whom do you prefer to seek help, a human therapist or an AI-powered chatbot, when you have mental health-related questions.}
    \label{fig:user_preference}
\end{figure*}

When asked about whom they prefer to seek help from when they have mental health-related questions, our participants expressed a strong preference towards human therapists (76\%) as shown in Figure~\ref{fig:user_preference}: Q2. Only about 16\% of participants preferred LLMs and human therapists equally, and 9\% of participants had a moderate preference towards LLMs. No participants had a strong preference towards LLMs. However, when asked about whether they would personally use LLMs to answer their mental health-related questions (a question that did not pose a choice between human and chatbot, but focused on the likelihood of using LLMs for answering mental health-related questions), more than 40\% of participants reported being likely to use LLMs (Figure~\ref{fig:user_preference}: Q1).

\begin{figure*}[ht]
\noindent
\centering
    \includegraphics[width=\linewidth]{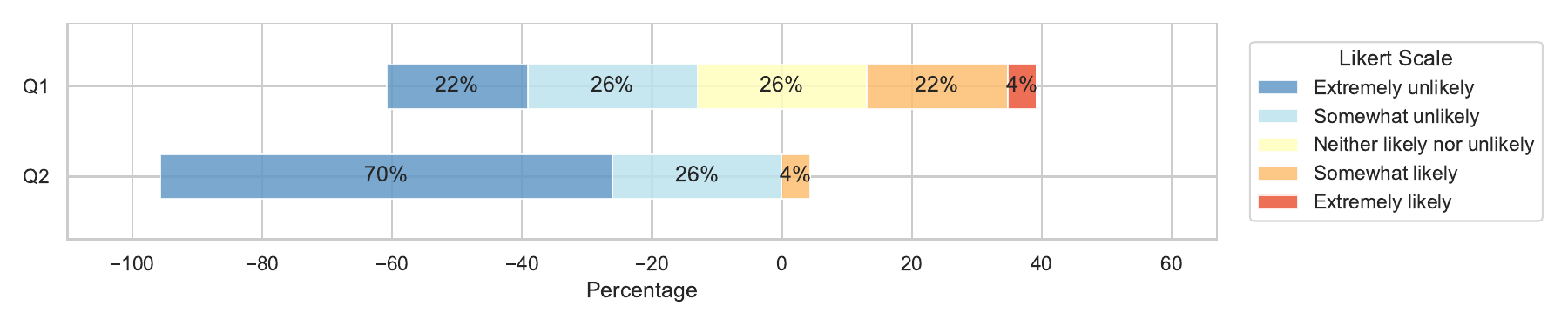}
    \vspace{-15pt}
    \caption{Distribution of user selection for Q1: As a mental health professional, would you recommend AI-powered chatbots to patients for obtaining general mental health information, and Q2: As a mental health professional, would you recommend AI-powered chatbots to patients for obtaining answers and advice similar to what is provided in a psychotherapy session.}
    \label{fig:therapist_preference}
\end{figure*}

For our licensed therapist participants, only approximately 25\% of them would recommend LLMs to patients for obtaining general mental health information, as shown in Figure~\ref{fig:therapist_preference}: Q1. LLMs are even less recommended for obtaining answers and advice similar to what is provided in a psychotherapy session, with only 4\% of therapist participants selecting ``Somewhat likely'' to recommend and 70\% of them selecting ``Extremely unlikely'' (Figure~\ref{fig:therapist_preference}: Q2). Overall, similar to user participants, our therapist participants did not express spreference towards using LLMs for mental health-related purposes, despite the higher ratings LLMs received.

\section{Discussion}

Our study set out to compare answers to mental health–related questions authored by LLMs and licensed therapists, asking both general users and therapists to evaluate them along qualities central to supportive communication. Across all rating dimensions, participants rated LLM-generated answers more highly than therapist answers. This pattern held for both general users and therapists, suggesting that contemporary models are capable of producing responses that are perceived as clearer, more encouraging, and more respectful than professional responses in this task context. At the same time, participants maintained a strong preference for seeking help from human therapists, and therapists expressed reluctance to recommend LLMs for anything beyond general information. This tension between perceived response quality and actual trust or acceptability mirrors prior work showing that AI systems can be persuasive and empathetic in style, while users remain cautious about entrusting them with sensitive domains~\cite{ayers2023comparing, guo2024large}. It also may indicate that while a LLM response may sound good, people are still unsure about the efficacy of responses from a LLM that could be used in therapeutic contexts.

Our results also revealed systematic variation across models. While ChatGPT and Gemini performed comparably, Llama was consistently rated highest on clarity, encouragement, respectfulness, and overall liking. This is notable given the common assumption that open-source models lag behind proprietary alternatives, suggesting instead that users may experience them as equally, if not more, effective at producing supportive responses. These results align with prior findings that stylistic fluency and surface-level coherence strongly shape user perceptions of chatbot quality, sometimes more than the underlying accuracy of content~\cite{bird2023generative, gabriel2024can}.

Demographic analysis indicated that participant background influenced perceptions in nuanced ways. Specifically, different therapy experiences shaped evaluations differently across models. While we expected that users with more experience in therapy might like or dislike LLM-generated answers, we found that those who had more than ten years of therapy experience rated Llama more positively. These patterns echo prior research showing that trust in AI is not uniform but contingent on users’ identities, prior experiences, and expectations~\cite{boyd2022usability, lee2020hear}. Importantly, we found that authorship judgments were above chance (45–60\% vs. 16\%), yet participants still experienced difficulty distinguishing therapist-written from LLM-generated responses. Notably, this difficulty did not appear to systematically affect ratings; participants who successfully identified LLM versus human responses nevertheless reported higher quality ratings for LLM answers. This suggests that the immediate qualities of a response, such as clarity, tone, and perceived empathy, may carry more weight than assumptions about its source, extending prior evidence that relational cues can outweigh knowledge of a system’s artificiality in shaping trust~\cite{darcy2021evidence, fitzpatrick2017delivering}. Some caution is warranted in interpreting this finding, however, as the authorship identification task was administered after participants completed their ratings. It remains possible that providing authorship information upfront could have altered how participants evaluated the responses.

Taken together, these findings highlight the promise and limits of deploying LLMs in mental health support. On the one hand, LLMs exhibit communicative competence that is comparable to professional responses, raising the possibility of supplementing scarce mental health resources with scalable digital tools. On the other hand, their lack of accountability, contextual judgment, and professional oversight constrains their acceptability as trusted sources of care, echoing long-standing arguments in HCI and digital health that ``competence'' does not equate to ``appropriateness'' in sensitive domains~\cite{ji2023rethinking, chung2023challenges}. Relatedly, communicative competence does not necessarily translate into therapeutic efficacy, highlighting an important area for future research.

As a final point, our team observed differences in the interpretation of results between members with psychology and computer science backgrounds. Our psychologists brought perspectives grounded in real-world experience with mental health issues, highlighting considerations that computer scientists alone might not identify. These observations underscore that interdisciplinary collaboration is crucial, and engaging with medical or mental health professionals is essential for ensuring that research on LLM-driven mental health technologies is both practical and applicable to real-world support systems.

\subsection{Legal and Privacy Challenges}
Despite LLM receiving higher ratings in our study, broader developments underscore the complexities and potential risks of deploying LLMs in mental health contexts. At a fundamental level, these models rarely meet the confidentiality and encryption standards required in therapeutic practice, raising concerns about surveillance, data sharing, and secondary use of sensitive disclosures~\cite{cabrera2023ethical, haque2023overview}. Without clear safeguards, users may be unaware of how their conversations are stored, used for training, or potentially exposed.

Furthermore, LLM-generated advice raises significant safety concerns. While these systems can produce responses that appear clear and empathetic, they may also generate inappropriate or harmful content in high-stakes situations. For example, the family of 16-year-old Adam Raine has filed a lawsuit alleging that prolonged interactions with ChatGPT validated Adam's distress and suicidal ideation~\cite{godoy2025openai}. They claimed that the system encouraged his self-harm, assisted with a suicide plan, and helped draft a suicide note. These concerns are also echoed by our licensed therapist participants. One participant worries that ``Chatbots can miss crucial information, including but not limited to, safety concerns and holds potential to invalidate the individual,'' while another commented that ``several of my current clients have had terrible experiences when they reached out in crisis and were naively connected with a chatbot.'' 

These concerns have already translated into legal action. In Illinois, the Wellness and Oversight for Psychological Resources (WOPR) Act now prohibits AI systems from independently delivering therapy, making diagnoses, or engaging directly with clients, unless supervised by a licensed professional. Similar laws have recently been passed in Nevada and Utah, and other states such as California, New Jersey, and Pennsylvania are actively considering setting forth restrictions of their own. These regulatory developments reflect a growing consensus across stakeholders from mental health professionals to legislators that while AI has the potential to provide mental health support, it should not be used or seen as a substitute for human therapists. Instead, LLMs’ application must be carefully scoped.

These developments align with HCI research emphasizing the need to design with boundaries, making explicit where AI systems can provide value and where human oversight is essential~\cite{denecke2021artificial}. At the same time, advances in privacy-preserving ML, such as federated learning and on-device processing, offer potential pathways to reconcile utility with confidentiality~\cite{ji2024mindguard}. For HCI, this underscores the need not only to evaluate user perceptions of LLMs, but also to consider how infrastructural choices about data storage and processing shape trust and adoption in practice.

\subsection{Design Implications}
The communicative competence of LLMs raises important design implications. Our findings confirm that LLMs can produce responses that feel empathetic and supportive, consistent with prior evidence that chatbots can foster disclosure and even therapeutic alliance~\cite{darcy2021evidence, fitzpatrick2017delivering}. Yet trust remained limited among both general users and therapists. This gap between perceived quality and willingness to rely on the system suggests that effective design cannot rely on surface-level fluency alone. Instead, designers must prioritize demonstrating efficacy, a gold standard of any treatment being added to healthcare. Additional concerns are transparency, accountability, and privacy to ensure that supportive-seeming responses do not obscure the absence of professional accountability. Prior research has shown that relational framing and self-disclosure cues can enhance trust and engagement, but in high-stakes contexts, such strategies risk encouraging over-attribution of competence to AI systems~\cite{ma2024understanding, gabriel2024can}. It would also be important to be able to audit these models to understand why they may give certain advice in different situations.

Design directions emerging from our findings emphasize the importance of scope, privacy, and escalation. LLMs should be positioned as sources of psychological education, journaling support, or triage rather than substitutes for professionals. Privacy-preserving architectures, such as local or federated deployments and ephemeral data storage, can help mitigate risks associated with sensitive disclosures. Finally, mechanisms for escalation to human professionals in moments of crisis are crucial, particularly as current systems may fail to escalate appropriately~\cite{heston2023evaluating}. These directions highlight the importance of designing with constraints: leveraging LLM strengths while acknowledging and addressing their limitations.

\subsection{Future Work}
Future studies should extend beyond perceptual evaluations and move toward longitudinal, in-situ investigations of how LLMs are integrated into everyday life. Mixed-methods approaches could examine how these systems shape trust, disclosure, and coping strategies over time, and whether they alter users’ pathways to professional care. Equally important are studies that probe how sensitive disclosures are processed and retained by LLMs, and how users perceive these risks, especially when prior work has shown that people’s privacy expectations are often context-dependent and shaped by design cues~\cite{haque2023overview}. Uncovering these expectations in mental health contexts will be critical among many concerns for informing safe deployments.

From a design perspective, future research should explore mechanisms such as ephemeral logging, explainable response generation, and selective disclosure architectures. These approaches may allow LLMs to provide supportive interaction while minimizing long-term exposure of sensitive information. Such explorations will help establish pathways for integrating LLMs into mental health–adjacent services, such as triage, self-help, or peer support, without undermining the confidentiality and contextual judgment that remain central to therapeutic practice. Future work might also investigate how LLMs could support human therapists in their clinical work, enhancing the quality of care they provide to clients.
\section{Conclusion}
This work offers one of the first systematic comparisons of therapist-written and LLM-generated responses to real mental health questions. Across clarity, respect, and supportiveness, LLM outputs were rated higher by both user participants and therapists, with Llama performing particularly well. Yet participants still preferred human therapists, and professionals were reluctant to recommend LLMs beyond general information, underscoring a gap between perceived quality and actual trust. We also found that experience undergoing therapy shaped participant evaluations, and that participants often struggled to distinguish authorship, though this did not affect their ratings. Together, these findings highlight both the promise and limits of LLMs, suggesting that LLMs can potentially extend access to supportive communication, but their lack of accountability, contextual judgment, and ethical safeguards prevents them from being acceptable substitutes for professional care. By situating LLMs as supplemental rather than substitutive resources, and by developing privacy-preserving and escalation-oriented architectures, researchers and designers can enable expanded access to supportive communication while safeguarding the values of trust, confidentiality, interpretability, and professional oversight that remain central to mental health care.

\bibliographystyle{ACM-Reference-Format}
\bibliography{ref}

\newpage
\appendix
\section{Model Specification}\label{sec:model_specification}
\begin{itemize}
    \item ChatGPT: chatgpt-4o-latest
    \item Gemini: gemini-1.5-pro
    \item Llama: 
Llama-3.1-70B-Instruct
\end{itemize}

\section{Survey Questions}\label{sec:survey}
\subsection{Example Questions}
Question (2) and (3), and question (5) and (6) are presented to participants in random order, so the participants might get therapist-written answer first or LLM-generated answer first. The rating scale for all questions is ``Strongly disagree'', ``Somewhat disagree'', ``Neutral'', ``Somewhat agree'', ``Strongly agree''.
\begin{enumerate}
    \item For each of the following questions, you will read a mental health-related question posted by a patient, and two different responses for the same question, which will appear one at a time. You will then rate the responses from different aspects.
    \item Patient Question with therapist-written answer
    \begin{itemize}
        \item (For therapists) This response is acceptable from a professional perspective.
        \item This response is clear. 
        \item This response answers the question. 
        \item This response is encouraging and supportive. 
        \item This response is respectful, accepting, and not judging. 
        \item Overall, I like this answer.
        \item Rating scale: Strongly disagree, Somewhat disagree, Neutral, Somewhat agree, Strongly agree
    \end{itemize}
    \item Patient Question with LLM-generated answer
    \begin{itemize}
        \item (For therapists) This response is acceptable from a professional perspective.
        \item This response is clear. 
        \item This response answers the question. 
        \item This response is encouraging and supportive. 
        \item This response is respectful, accepting, and not judging. 
        \item Overall, I like this answer.
    \end{itemize}
    \item You will now read the same questions and responses again. This time, carefully rate how much you think each response is likely generated by an AI-powered chatbot.
    \item Patient Question with therapist-written answer
    \begin{itemize}
        \item This response is generated by an AI.
    \end{itemize}
    \item Patient Question with LLM-generated answer
    \begin{itemize}
        \item This response is generated by an AI.
    \end{itemize}
\end{enumerate}

\subsection{Demographics and Additional Information}
\begin{enumerate}
    \item (For users)  Would you personally use AI-powered chatbots to answer your mental health-related questions?
    \item (For users) From whom do you prefer to seek help, a human therapist or an AI-powered chatbot, when you have mental health-related questions?
    \item (For therapists) As a mental health professional, would you recommend AI-powered chatbots to patients for obtaining general mental health information?
    \item (For therapists) As a mental health professional, would you recommend AI-powered chatbots to patients for obtaining answers and advice similar to what is provided in a psychotherapy session?
    \item Do you have any opinion on using chatbots for mental health related purposes that you would like to share with us?
    \item (For users) Have you ever received services from a mental health professional, such as a psychologist, counselor, therapist, or psychiatrist? If so, for how long?
    \item (For therapists) How long have you been a therapist for?
    \item What is your age?
    \item What gender do you identify with?
    \item Which group(s) do you identify with (select all that apply)?
    \item Do you live in the U.S.?
    \item Which state do you live in?
    \item What is your highest level of education attained?
\end{enumerate}

\section{CLMM Specification}\label{sec:statistics}

We model the ordinal response $D_{ij} \in \{-4,-3,\dots,4\}$, defined as the
rating difference between large language model (LLM) outputs and human responses, 
using a cumulative link mixed model (CLMM) with a logit link. 
Let $i$ index participants and $j$ index treatments ($\text{gpt}$, $\text{gemini}$, $\text{llama}$). 
The cumulative probability that participant $i$ under treatment $j$ rates at or below category $k$ is

\begin{equation}
\Pr(D_{ij} \leq k \mid \mathbf{x}_{ij}, u_i) 
= \text{logit}^{-1}\!\left(\theta_k - \eta_{ij} - u_i\right), 
\quad k = -4,\dots,3,
\end{equation}

where $\theta_k$ are threshold (cutpoint) parameters, and 
$u_i \sim \mathcal{N}(0,\sigma^2)$ is a participant-specific random intercept.

The linear predictor $\eta_{ij}$ is specified as
\begin{align*}
\eta_{ij} = 
\alpha_{\text{gemini}} I(\text{Treatment} = \text{gemini})
+ \alpha_{\text{llama}} I(\text{Treatment} = \text{llama}) \\
+ \sum_m \beta_m X_{ijm}
+ \sum_\ell \gamma_\ell \big(\text{Treatment} \times \text{Experience}\big)_{ij\ell},
\end{align*}
where $\text{Experience}$ refers to therapy experience, 
$X_{ijm}$ are demographic covariates (education, age, race, state, gender, and 
the indicator \texttt{can\_distinguish}), and 
the interaction terms capture moderation by therapy experience.

\end{document}